\documentclass{aastex63}

\received{October 28, 2019}
\revised{January 03, 2020}
\accepted{February 19, 2020}
\submitjournal{ApJ}

\shorttitle{Sunspot group decay}
\shortauthors{J. Murak\"ozy}

\begin{document}

\title{Study of the decay rates of the umbral area of sunspot groups by using a high resolution database}

\correspondingauthor{Judit Murak\"ozy}
\email{murakozy.judit@csfk.mta.hu}

\author[0000-0001-6920-259X]{Judit Murak\"ozy}
\affiliation{Konkoly Thege Mikl\'os  Astronomical Institute\\ Research Centre for Astronomy and Earth Sciences\\
Konkoly-Thege Mikl\'os \'ut 15-17.\\
Budapest, 1121, Hungary}

\begin{abstract}

The emergence and decay of the sunspot groups are important components of the solar dynamo models. There are two different types of studies on the evolution of active regions. One of them is based on fewer data with higher spatial resolution, the other one uses more data with lower spatial resolution. The input data of the present study allow the investigation with high-resolution both spatially and temporally. The temporal resolution of the SoHO Debrecen Sunspot Database (SDD) is one and a half hours, and it also makes possible to identify all individual sunspots with the position, area, and magnetic polarity. More than 200 sunspot groups have been selected, which have clear maxima on the solar disc, and the decrease of their umbral area is observable during at least four days. The decay rates were calculated by using two data: the umbral area and the number of contained sunspots -- these decay rates were computed for the total umbral area of sunspot groups and their leading and following parts. The decay rate has a linear area dependency, and it is higher for the following part than for the leading one.
\end{abstract}


\section{Introduction} \label{sec:intro}

The evolution of solar active regions has been investigated in several papers by different methods. The most thoroughly investigated phenomenon is the sunspot's decay and its properties. Some papers use large samples of sunspots while others examine a single selected spot or make numerical simulations. \citet{2018ApJ...865...88C} use 196 sunspot drawings made under the Maunder Minimum. They examined 48 different sunspots by using their umbra--penumbra ratio and found that sunspots with higher U/P values show faster decay. \citet{2008SoPh..250..269H} investigated sunspot groups, and spots focused on their latitudinal distributions, area, cycle, and cycle phase dependencies as well.
\citet{2018A&A...620A.191B} have pointed out linear decay of the umbral vertical magnetic field and area as well as increasing penumbral area by studying only one decaying sunspot. 
\citet{2007ApJ...671.1013D} studied a decaying follower sunspot observed for six days and found that magnetic flux removed from the spot.
The results of the modeled decay show good agreement with the empirical results.
\citet{2015ApJ...814..125R} has calculated the decay rate of the naked spots by using numerical simulations and obtained 10$^{21}$ Mx/day. \citet{2015ApJ...800..130L} obtained a parabolic decay law for the sunspot area, and their results agree with the predictions of \citet{1997ApJ...485..398P}.
\citet{1997ASPC..118..145P} and \citet{1997SoPh..176..249P} published the most similar work to the present study examining individual sunspots.
They also used a large sample and pointed out that the decay rate is faster in the earlier phase of the decay than in the later phase. \citet{1999SoPh..188..315P} showed that the neighborhood of the sunspots also contributes to the decay rate of sunspots.

Most of the papers are dealing with the decay of individual sunspots while the present investigation focuses only on the decay of sunspot groups, which is a less analyzed process.
The papers on this topic use either large sample without distinction to the magnetic polarity or distinguish the leading and following parts on a small sample. There also are some studies based on a small sample. \citet{1996JKAS...29....9L} investigated the growth and decay of sunspot
groups on two different groups. There are also theoretical studies of the decay. 
\citet{1975SoPh...42..107K} investigated the sunspot group's decay by using a simple theoretical model and compared their results to the results of \citet{1963BAICz..14...91B} based on observations. They found that the long-lived groups decay much slower, while in the first days of the growth or decay, the area of groups varies faster.

The earlier works using large samples are based on sunspot catalogs which do not contain magnetic polarity data.
\citet{2012Ap&SS.338..217J} and \citet{1995SoPh..157..389L} used the Greenwich Photoheliographic Results (GPR), and \citet{1993SoPh..147....1H} used Mount Wilson sunspot data and obtained results on the decay of the groups and plages.

If the authors wanted to distinguish between the opposite polarities, they had to process space-borne observations which necessarily restricted the size of the sample as in the paper of \citet{2017ApJ...842....3N}. They studied the growth and decay rates in 10 active regions by using SDO/HMI continuum intensity and vector magnetogram data. Their main aim was the study of the leading and following parts' development, but they obtained results for the decay as well.

\section{Data and methods}
\subsection{Database with high spatial and temporal resolutions}
\label{databases}
The input data of the present work are taken from the Debrecen sunspot databases \citep{2016SoPh..291.3081B, 2017MNRAS.465.1259G}, the only catalogues containing the position and area data of all sunspot groups and each individual sunspot (umbra and penumbra). The early member of this set is the Debrecen Photoheliographic Data (DPD) reduced from ground based observations. Later members are the SDD (SOHO\/MDI-Debrecen Sunspot Data)  and HMIDD (SDO/HMI-Debrecen Sunspot Data), these used both images and  magnetograms of the space instruments and they give also the magnetic fields of the sunspots. The detailed description of these databases is published by \citep{2016SoPh..291.3081B}. The technique of the sunspot data production is described by \citep{1998SoPh..180..109G, 2012SoPh..280..365G}. The measurement of the areas of a sunspot group is illustrated by \citet{2015SoPh..290.1627G} in his Figure 1. 

The present work uses the SDD containing sunspot data from 1996 until 2010, in particular the area data of sunspot umbrae corrected for foreshortening. We use the total umbral area of the entire active region (hereafter A$_g$) and that of the leading (A$_l$) and following (A$_f$) parts separately. Due to the 1.5 hourly temporal resolution of the data set the evolution of the sunspot groups can be studied in detail. The magnetic field data of sunspot umbrae make possible to separate the leading and following parts of the sunspot groups.

\begin{figure}[h!]
    \includegraphics[width=0.8\textwidth, angle=0]{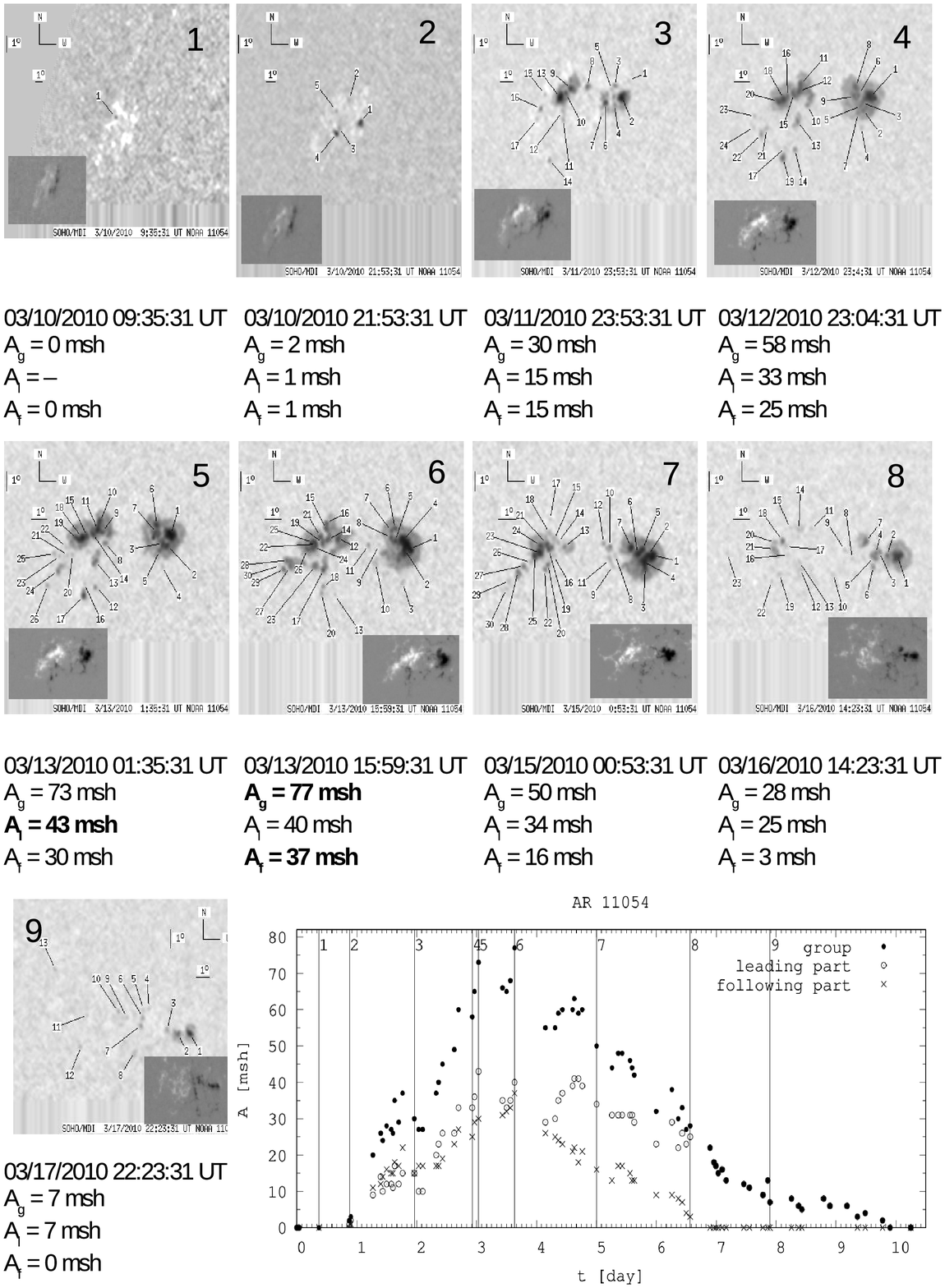}
    \caption{Evolution of active region 11054. The right bottom figure shows the time-series of the total umbral area of sunspot groups (depicted by dots) and their leading (marked by empty dots) and following parts (marked by crosses) of this AR. The numbered vertical lines mark the times of the chose observations. The time starts from the first observation of the AR. The numbered panels are AR's images cutted out from the SoHO spacecraft white light observations. For the shake of the easy comparison there is a smaller gray box on every white light picture. These are the part of magnetograms coupled to the AR's. Below the observations one can find the umbral area of the group (A$_g$), the leading part (A$_l$) and the following part (A$_f$) at the time of the observation. Their maximum values are marked by boldface. The area data are taken from the numerical SDD catalog, while the images are parts of SoHO white-light observations, they are accessible among the on-line tools of the catalog (available on-line at http://fenyi.solarobs.csfk.mta.hu/SDD). \label{fig:samplear}}
\end{figure}

\subsection{Observed decay of sunspot groups}
\label{secsec}
The following selection criteria were applied. The observable growth and decay extend at least to 40\% of the maximum total area and the decay lasts at least four days. The sunspot groups are taken into account within 65$^{\circ}$ from the central meridian. The maximum area state of the sunspot groups can be regarded as their equilibrium state.\\
Each selected sunspot group has also been checked visually. This ensures that their area decays unambiguously and the leading or following parts are not developing or unchanged while the total area is decreasing. By using the described selection method, 206 appropriate groups can be chosen from the SDD catalog over the time-span of 1996 -- 2010. 

The decay rates are calculated as the ratio between the difference of areas measured in two observations and the time interval between them. This decay rate was calculated for (the total umbral area of sunspot group) and its leading and following parts as well. It is also necessary to treat the leading and following parts separately because their evolutionary curves may have a phase difference. In the right bottom panel of Fig.~\ref{fig:samplear} and on panels 5-6 one can see, while the following part of AR 11054 is still growing, its leading part is already decaying.
Otherwise, AR 11054 can be considered as a typical sunspot group since the following part reaches its maximum state later than the leading one and its maximum is lower than that of the leading one, as in most cases \citep{2014SoPh..289..563M}.

The decay rates ($d$) were computed by two different methods. One  of them is a simple method: the last observed and the maximum data were taken into account in the following way:

\begin{equation}
    d=\frac{A_{fin}-A_{max}}{t_{fin}-t_{max}}.
	\label{eq:decay}
\end{equation}
where $A_{fin}$ is the last observed umbral area at the last $t_{fin}$ time, while $A_{max}$ is the equilibrium area at the time of the maximum ($t_{max}$).

By the other method, the decay rates have been determined by using the daily changes of the sunspot groups from their maximum state until their finally observed state. The average value of these daily changes gives the decay rate of the sunspot group. The first data of the groups have been taken into account as the daily values every day. Although this method results in higher deviations due to the fluctuations of the daily values, it gives similar results to the first method.

\subsection{Analytically computed decay rates}
The third method to determine the decay rates is as follows.
The following asymmetric Gaussian function has been fitted to the time series of the sunspot group's area data
\begin{equation}
    f(x)=H\cdot exp-\frac{(t-t_M)^2}{w\cdot(1+a(t-t_M))}
	\label{eq:asymmgaussian}
\end{equation}
\noindent
where $H$ and t$_M$ are the height and time of the maximum of the time-series while $w$ and $a$ determine the width and the asymmetry, respectively. This function has been used for the development study by \citet{2014SoPh..289..563M}. The fit has been done for the entire selected sample. 
The decay rate was determined as the steepness of function (\ref{eq:asymmgaussian}) at the half-maximum level in the decreasing phase. These decay rates (Fig.~\ref{fig:samplear}) are also computed separately for the total umbral area of sunspot groups ($s_g$) as well as their leading ($s_l$) and following parts ($s_f$).

\section{Results in the study of sunspot group decay rates}
\label{decrate}
The decay of the active region magnetic fields is studied by tracking the temporal variation of the total umbral area of the sunspot group which represents best the evolution of the emerged magnetic field.

\begin{figure}[ht!]
 \includegraphics[width=0.7\textwidth, angle=0]{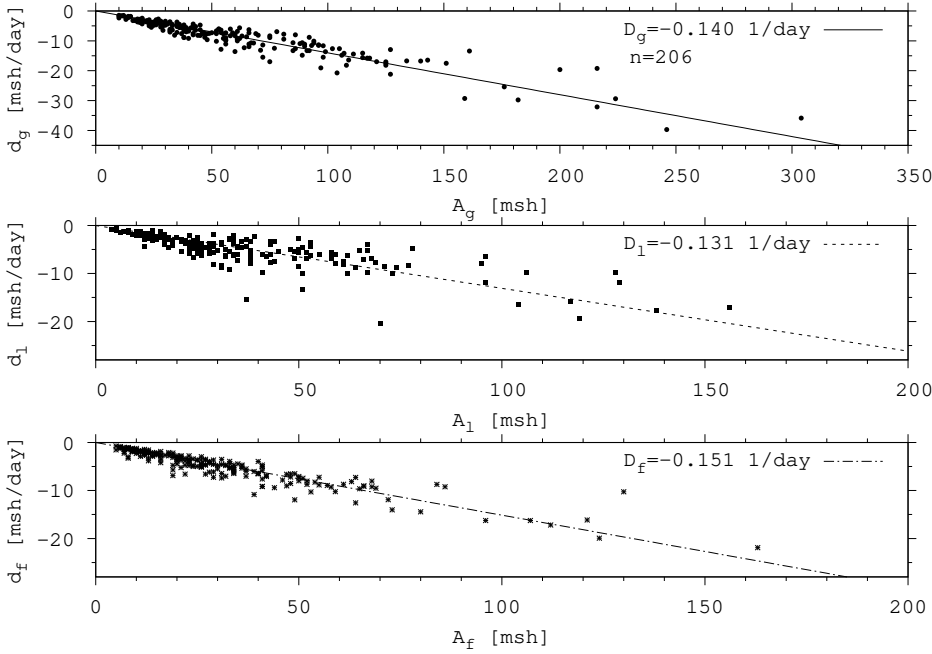}
 \caption {Equilibrium area dependence of the decay rate of sunspot groups and their opposite polarity parts by using Eq.~\ref{eq:decay}. The top panel shows the decay rate of the sunspot groups while the middle and the bottom panels show that for the leading and following parts. The total number of the studied groups is 206, and the values of the steepness of the fitted functions can be seen in the proper panels.  \label{fig:areadecaysimple}}
\end{figure}

\begin{figure}[hb!]
 \includegraphics[width=0.7\textwidth, angle=0]{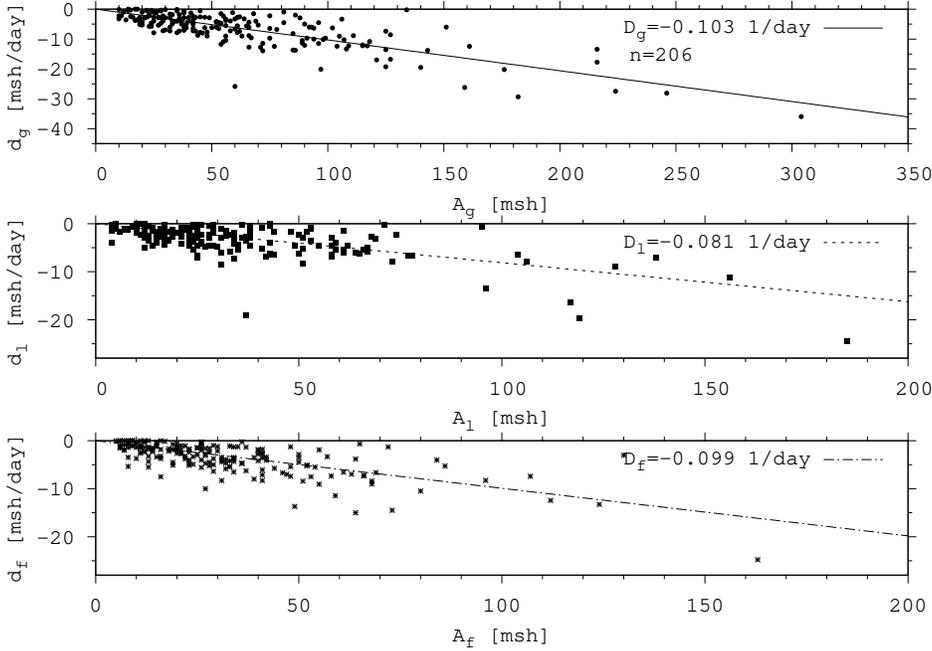}
  \caption{Total area dependence of the decay rate of sunspot groups and their opposite polarity parts by using the average daily decay method. The top panel shows the decay rate of the sunspot groups while the middle and the bottom panels show that for the leading and the following parts. The total number of studied groups is 206. The values of the steepness of the fitted functions can be seen in the proper panels. \label{fig:areadecayavg}}
\end{figure}

The decay rates computed by the two methods described in section \ref{secsec} have been determined for both the area and sunspot number. 
Figs.~\ref{fig:areadecaysimple} and \ref{fig:areadecayavg} show the area dependency of the areal decay rates. The areal decay rates depend linearly on the area of the equilibrium state of the  total umbral area of group and also of its leading and following parts. The rates of the areal decay are higher in the following parts by all methods.

\begin{figure}[h!]
 \includegraphics[width=0.7\textwidth, angle=0]{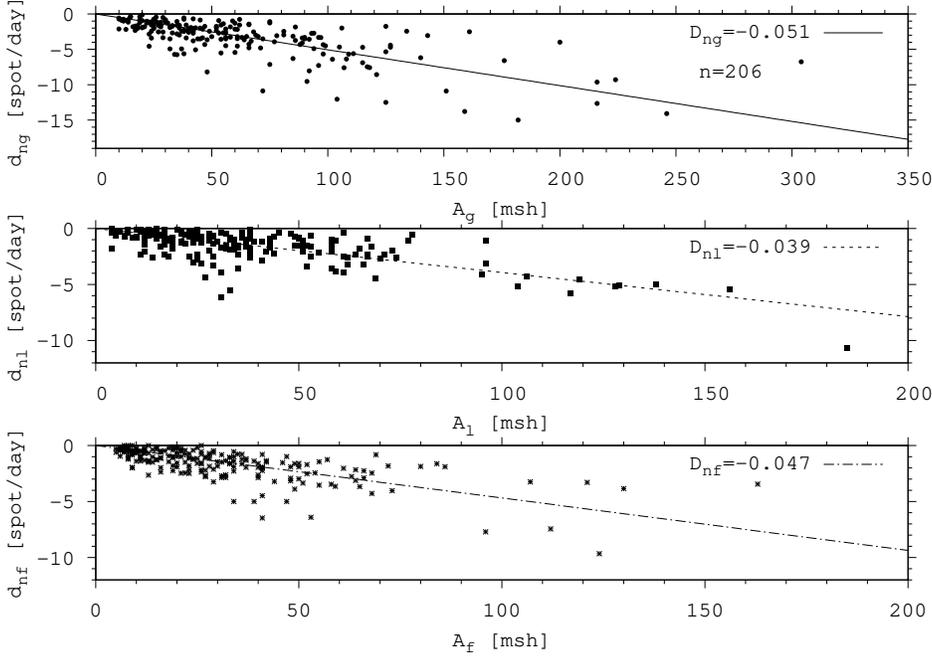}
    \caption{The decay rate of the sunspot number dependency on the area of the equilibrium state of proper parts. The decay rate of the total umbral area of groups (top panel), their leading parts (middle panel), and following parts (bottom panel). The total number of studied groups is 206. The lines depict the linear dependency on the proper maximum area. The steepness parameter of the fitted linear functions can be seen in the appropriate panels. \label{fig:numberdecay}}
\end{figure}

\begin{figure}[h!]
	 \includegraphics[width=0.7\textwidth, angle=0]{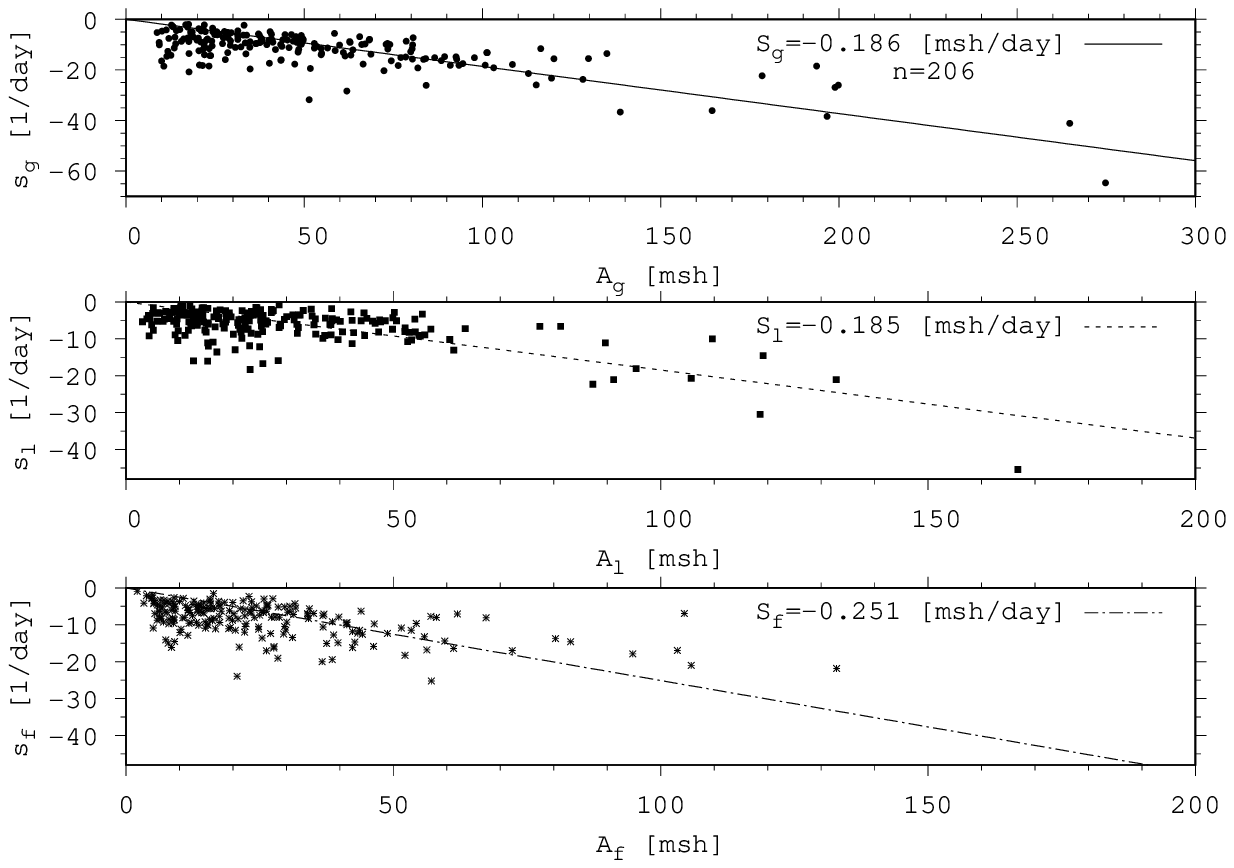}
    \caption{The decay rate of the sunspot area dependency on the area of the equilibrium state of proper parts. The decay rates calculated by using the fitted asymmetric Gaussian function. Decay rates of the total umbra area of groups (top panel), their leading parts (middle panel), and following parts (bottom panel). The number of studied groups is 206. The lines depict the linear dependency on the maximum area. The equations of the fitted functions can be seen in the proper panels. \label{fig:fitted}}
\end{figure}

Since the results obtained by the two different methods are similar not only for the areal decay but also for the number decay therefore only one figure is presented here that shows the number decay obtained by the simple method (Eq.~\ref{eq:decay}). 
Fig.~\ref{fig:numberdecay} shows that the regression rate of the sunspot number also depends on the maximum area, and this rate is lower for the leading part than for the following one. It means that while the following part usually has a higher number of sunspots than the leading part, it disappears more rapidly. At the end of the first and steepest part of the decay process, there will be only one large remaining leading spot without any following one, or in some cases, there may be some tiny spots with the following polarity.

The decay rates computed by the method of asymmetric Gauss function (Equation~\ref{eq:asymmgaussian}) are shown in Fig.~\ref{fig:fitted}. The top, middle, and bottom panels pertain to the total umbral area of groups, and their leading and following parts, respectively.

The decay rate can be determine as the slope of the asymmetric Gaussian function at the half-height of the curve. The analytically determined decay rate is a more accurate and reliable procedure than the two other ones because it does not need the observations of the last states. Nevertheless, the results given by this procedure are similar to those of the previous methods. The decay rate has a total area dependence, and there is an asymmetry between the leading and following decay rates, i.e., the decay rate of the following parts is higher than the leading one.

\begin{figure}[h!]
 \includegraphics[width=0.5\textwidth, angle=-90]{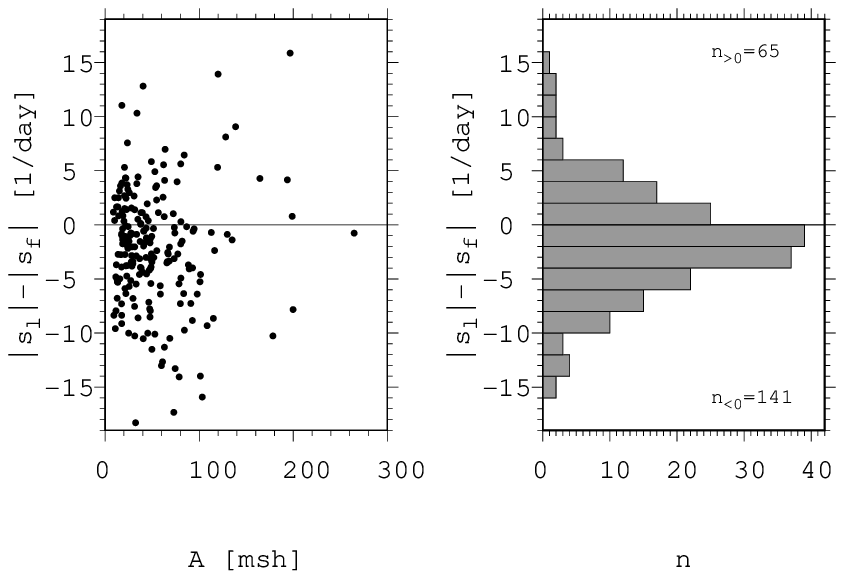}
    \caption{Area dependency of the difference between the magnitude of leading and following decay steepness (left panel). The case number counted in two msh/day bins based on the left panel (right panel). \label{fig:asymm}}
\end{figure}

\begin{figure}[ht!]
 \includegraphics[width=0.6\textwidth, angle=0]{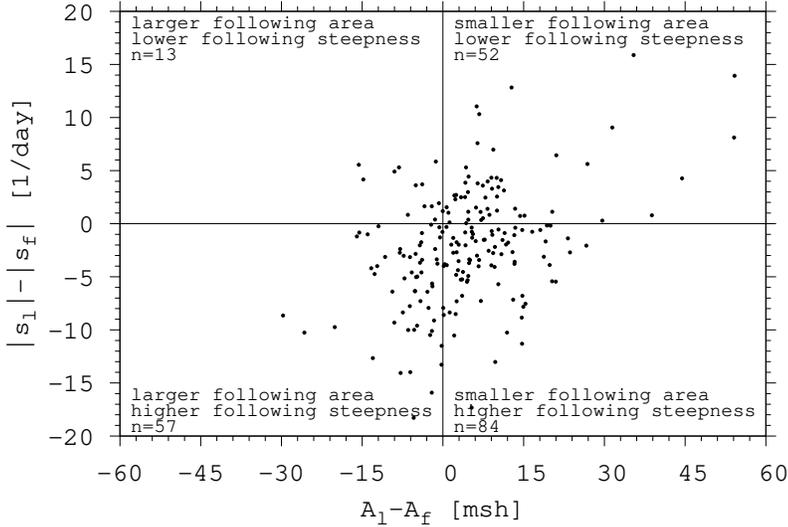}
    \caption{Relationship between asymmetries of leading--following areas and steepness of the asymmetric Gaussian functions fitted to the time-series of the decay of both parts. \label{fig:asymmasymm}}
\end{figure}
\pagebreak

The asymmetry between the leading--following decay rates is characterized by the difference between the steepnesses of the fitted Gaussian functions (Fig.~\ref{fig:asymm}). It can be seen that the $|s_l|-|s_f|$ difference is independent of the AR's total area. The majority of the cases (68\%) are negative, meaning that the decay of the following part is steeper than that of the leading part.

\begin{figure}[ht]
	 \includegraphics[width=0.95\textwidth, angle=0]{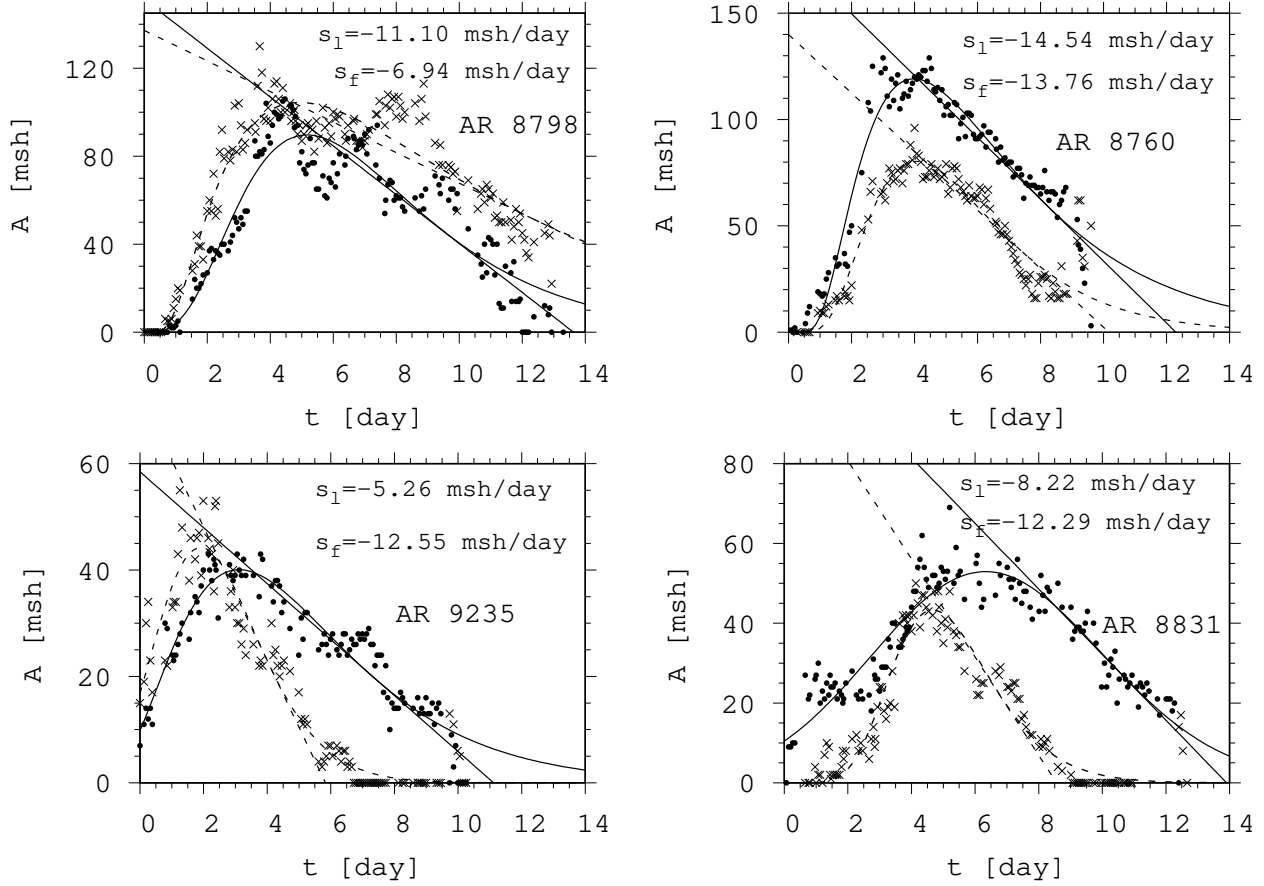}
    \caption{Examples of evolutionary courses for the four quadrants of Fig.~\ref{fig:asymmasymm}. Top panels: higher leading steepness with a larger following area (on the left) and larger leading area (on the right). Bottom panels: higher following steepness with a larger following area (on the left) and larger leading area (on the right). The leading and following areas of the AR have been depicted by dots and crosses, respectively. These time series fitted by the asymmetric Gaussian function (Eq.~\ref{eq:asymmgaussian}). The slope of this function has been fitted by a linear function, which gives us the steepness of the leading part decay (continuous curve) and the following part decay (dashed curve). \label{fig:samples}}
\end{figure}

These results apply for the majority of cases. Fig.~\ref{fig:asymmasymm} shows four different cases of sizes and decay rates of leading and following parts. It can be seen that the most populated quadrant is the right lower one; this contains the cases of the smaller following area and higher following decay steepness; this is the most typical case. However, the rest of the combinations, the right upper and left lower quadrants are also not negligible. The least typical case is when the area of the leading part is smaller, and its decay rate is higher than those of the following part. These four combinations are illustrated with four examples in Fig.~\ref{fig:samples}.

\begin{deluxetable*}{cccccc}[h!]
\tablenum{1}
\tablecaption{Decay rates by using various methods}\label{tab:results}
\tablewidth{0pt}
\tablehead{
 parts & &\colhead{U area decay} && \colhead{Sunspot number decay} \\ 
 & \colhead{simple method} & \colhead{average method} & \colhead{Gaussian fit} & \colhead{simple method}}
\startdata
group  & -0.140$\pm$0.002 & -0.103$\pm$0.004 & -0.186$\pm$0.006 & -0.051$\pm$0.002 \\
leading part & -0.131$\pm$0.004 & -0.081$\pm$0.005 & -0.185$\pm$0.008 & -0.039$\pm$0.002 \\
following part & -0.151$\pm$0.003 & -0.099$\pm$0.005 & -0.251$\pm$0.011 & -0.047$\pm$0.002
\enddata
\end{deluxetable*}

The data of Table~\ref{tab:results} are numerical corroborations of the previously detailed leading--following asymmetry of the sunspot groups' decay. In the case of big and old sunspot groups, it can be observed that the sporadic follower spots disappear sooner, while the bigger leading spot usually remains visible on the disc for a longer time.

\section{Conclusions}

Two hundred six bipolar sunspot groups have been examined. They were selected by scripts from the SDD database, but each of them has been checked visually by using the image platform of the Debrecen databases. The decay rates have been determined for the total umbral area of the selected ARs as well as their leading and following parts. Three methods applied for the area decay and two methods for the sunspot number. 
The most reliable method is the determination of the steepness because it only takes into account the real declining phase of the fitted asymmetric Gaussian function and omits the limb with almost zero areas.

The following results can be listed:

1. The decay rate of the total umbral area of a sunspot group depends on the total umbral area at its maximum state. The larger is the maximum area, the more rapid is the decay. This result has been obtained by three methods.

This result is similar to that of \citet{2014SoPh..289...91G}. They investigated a considerable sample but did not make any distinction between the leading and following parts of sunspot groups.

2. The decay rate of the following part is typically higher than that of the leading part. This result is also independent of the used calculation method. 

3. The decrease of the sunspot number in the sunspot group exhibits the same dependencies as the area decays.

These results are consistent with those of \citet{2017ApJ...842....3N}; they studied the process of the sunspot group evolution on ten selected groups and studied mainly the growth. They studied the decay on a part of their sample. They also find that the following part decays more rapidly than the leading part, and the growth rate is higher than the decay rate.

The results of the present paper, together with those of the previous paper of \citet{2014SoPh..289..563M}, provide a detailed description of the dynamics of emergence and decay of sunspot groups on a large sample thanks to the most detailed Debrecen sunspot databases.

\section*{Acknowledgements}

The research leading to these results has received funding from National Research, Development and Innovation Office -- NKFIH, 129137. Thanks are due to Andr\'as Ludm\'any for reading the manuscript.
\nopagebreak




\bibliography{Murakozy_apj_2019}{}

\begin{thebibliography}{}
\expandafter\ifx\csname natexlab\endcsname\relax\def\natexlab#1{#1}\fi
\providecommand{\url}[1]{\href{#1}{#1}}
\providecommand{\dodoi}[1]{doi:~\href{http://doi.org/#1}{\nolinkurl{#1}}}
\providecommand{\doeprint}[1]{\href{http://ascl.net/#1}{\nolinkurl{http://ascl.net/#1}}}
\providecommand{\doarXiv}[1]{\href{https://arxiv.org/abs/#1}{\nolinkurl{https://arxiv.org/abs/#1}}}

\bibitem[{{Baranyi} {et~al.}(2016){Baranyi}, {Gy{\H{o}}ri}, \&
  {Ludm{\'a}ny}}]{2016SoPh..291.3081B}
{Baranyi}, T., {Gy{\H{o}}ri}, L., \& {Ludm{\'a}ny}, A. 2016, \solphys, 291,
  3081, \dodoi{10.1007/s11207-016-0930-1}

\bibitem[{{Benko} {et~al.}(2018){Benko}, {Gonz{\'a}lez Manrique}, {Balthasar},
  {G{\"o}m{\"o}ry}, {Kuckein}, \& {Jur{\v{c}}{\'a}k}}]{2018A&A...620A.191B}
{Benko}, M., {Gonz{\'a}lez Manrique}, S.~J., {Balthasar}, H., {et~al.} 2018,
  \aap, 620, A191, \dodoi{10.1051/0004-6361/201834296}

\bibitem[{{Bumba}(1963)}]{1963BAICz..14...91B}
{Bumba}, V. 1963, Bulletin of the Astronomical Institutes of Czechoslovakia,
  14, 91

\bibitem[{{Carrasco} {et~al.}(2018){Carrasco}, {Garc{\'\i}a-Romero}, {Vaquero},
  {Rodr{\'\i}guez}, {Foukal}, {Gallego}, \&
  {Lef{\`e}vre}}]{2018ApJ...865...88C}
{Carrasco}, V.~M.~S., {Garc{\'\i}a-Romero}, J.~M., {Vaquero}, J.~M., {et~al.}
  2018, \apj, 865, 88, \dodoi{10.3847/1538-4357/aad9f6}

\bibitem[{{Deng} {et~al.}(2007){Deng}, {Choudhary}, {Tritschler}, {Denker},
  {Liu}, \& {Wang}}]{2007ApJ...671.1013D}
{Deng}, N., {Choudhary}, D.~P., {Tritschler}, A., {et~al.} 2007, \apj, 671,
  1013, \dodoi{10.1086/523102}

\bibitem[{{G{\'o}mez} {et~al.}(2014){G{\'o}mez}, {Curto}, \&
  {Gras}}]{2014SoPh..289...91G}
{G{\'o}mez}, A., {Curto}, J.~J., \& {Gras}, C. 2014, \solphys, 289, 91,
  \dodoi{10.1007/s11207-013-0323-7}

\bibitem[{{Gy{\H{o}}ri}(1998)}]{1998SoPh..180..109G}
{Gy{\H{o}}ri}, L. 1998, \solphys, 180, 109, \dodoi{10.1023/A:1005081621268}

\bibitem[{{Gy{\H{o}}ri}(2012)}]{2012SoPh..280..365G}
---. 2012, \solphys, 280, 365, \dodoi{10.1007/s11207-012-9987-7}

\bibitem[{{Gy{\H{o}}ri}(2015)}]{2015SoPh..290.1627G}
---. 2015, \solphys, 290, 1627, \dodoi{10.1007/s11207-015-0714-z}

\bibitem[{{Gy{\H{o}}ri} {et~al.}(2017){Gy{\H{o}}ri}, {Ludm{\'a}ny}, \&
  {Baranyi}}]{2017MNRAS.465.1259G}
{Gy{\H{o}}ri}, L., {Ludm{\'a}ny}, A., \& {Baranyi}, T. 2017, \mnras, 465, 1259,
  \dodoi{10.1093/mnras/stw2667}

\bibitem[{{Hathaway} \& {Choudhary}(2008)}]{2008SoPh..250..269H}
{Hathaway}, D.~H., \& {Choudhary}, D.~P. 2008, \solphys, 250, 269,
  \dodoi{10.1007/s11207-008-9226-4}

\bibitem[{{Howard}(1993)}]{1993SoPh..147....1H}
{Howard}, R.~F. 1993, \solphys, 147, 1, \dodoi{10.1007/BF00675483}

\bibitem[{{Javaraiah}(2012)}]{2012Ap&SS.338..217J}
{Javaraiah}, J. 2012, \apss, 338, 217, \dodoi{10.1007/s10509-011-0932-2}

\bibitem[{{Krause} \& {Ruediger}(1975)}]{1975SoPh...42..107K}
{Krause}, F., \& {Ruediger}, G. 1975, \solphys, 42, 107,
  \dodoi{10.1007/BF00153288}

\bibitem[{{Lee} {et~al.}(1996){Lee}, {Yun}, {Moon}, \&
  {Wang}}]{1996JKAS...29....9L}
{Lee}, S.~W., {Yun}, H.~S., {Moon}, Y.~J., \& {Wang}, J.~L. 1996, Journal of
  Korean Astronomical Society, 29, 9

\bibitem[{{Litvinenko} \& {Wheatland}(2015)}]{2015ApJ...800..130L}
{Litvinenko}, Y.~E., \& {Wheatland}, M.~S. 2015, \apj, 800, 130,
  \dodoi{10.1088/0004-637X/800/2/130}

\bibitem[{{Lustig} \& {Wohl}(1995)}]{1995SoPh..157..389L}
{Lustig}, G., \& {Wohl}, H. 1995, \solphys, 157, 389,
  \dodoi{10.1007/BF00680629}

\bibitem[{{Murak{\"o}zy} {et~al.}(2014){Murak{\"o}zy}, {Baranyi}, \&
  {Ludm{\'a}ny}}]{2014SoPh..289..563M}
{Murak{\"o}zy}, J., {Baranyi}, T., \& {Ludm{\'a}ny}, A. 2014, \solphys, 289,
  563, \dodoi{10.1007/s11207-013-0416-3}

\bibitem[{{Norton} {et~al.}(2017){Norton}, {Jones}, {Linton}, \&
  {Leake}}]{2017ApJ...842....3N}
{Norton}, A.~A., {Jones}, E.~H., {Linton}, M.~G., \& {Leake}, J.~E. 2017, \apj,
  842, 3, \dodoi{10.3847/1538-4357/aa7052}

\bibitem[{{Petrovay} {et~al.}(1999){Petrovay}, {Mart{\'\i}nez Pillet}, \& {van
  Driel-Gesztelyi}}]{1999SoPh..188..315P}
{Petrovay}, K., {Mart{\'\i}nez Pillet}, V., \& {van Driel-Gesztelyi}, L. 1999,
  \solphys, 188, 315, \dodoi{10.1023/A:1005213212336}

\bibitem[{{Petrovay} \& {Moreno-Insertis}(1997)}]{1997ApJ...485..398P}
{Petrovay}, K., \& {Moreno-Insertis}, F. 1997, \apj, 485, 398,
  \dodoi{10.1086/304404}

\bibitem[{{Petrovay} \& {van
  Driel-Gesztelyi}(1997{\natexlab{a}})}]{1997ASPC..118..145P}
{Petrovay}, K., \& {van Driel-Gesztelyi}, L. 1997{\natexlab{a}}, in
  Astronomical Society of the Pacific Conference Series, Vol. 118, 1st Advances
  in Solar Physics Euroconference. Advances in Physics of Sunspots, ed.
  B.~{Schmieder}, J.~C. {del Toro Iniesta}, \& M.~{Vazquez}, 145

\bibitem[{{Petrovay} \& {van
  Driel-Gesztelyi}(1997{\natexlab{b}})}]{1997SoPh..176..249P}
{Petrovay}, K., \& {van Driel-Gesztelyi}, L. 1997{\natexlab{b}}, \solphys, 176,
  249, \dodoi{10.1023/A:1004988123265}

\bibitem[{{Rempel}(2015)}]{2015ApJ...814..125R}
{Rempel}, M. 2015, \apj, 814, 125, \dodoi{10.1088/0004-637X/814/2/125}

\end{thebibliography}
\bibliographystyle{aasjournal}

\end{document}